\documentclass[12pt]{article}
\usepackage{amsmath}
\usepackage{float}
\usepackage{threeparttable}
\usepackage{array}
\usepackage{indentfirst}
\usepackage{booktabs}
\usepackage{caption}
\usepackage{graphicx, subfig}
\usepackage{color}
\usepackage{cite}
\usepackage{epsfig}
\usepackage{times}
\usepackage{hyperref}
\hypersetup{
	colorlinks=true,
	linkcolor=blue,
	urlcolor=blue,
	citecolor=red,
	linktoc=all
}

\makeatletter
\def\@biblabel#1{(#1)}
\def\@cite#1#2{[#1\if@tempswa , #2\fi]}
\makeatother

\begin{document}
	\thispagestyle{plain}
	
	\title{\bf Arbitrariness and Usefulness of the Expressions of Elastic wave's Energy, Momentum and Angular Momentum}
	\author{\normalsize Zi-Wei Chen, Bang-Hui Hua and Xiang-Song Chen\thanks{Email:
			cxs@hust.edu.cn}}
	\date{}
	\maketitle
~~\noindent{\small School of Physics, Huazhong University of Science
	and Technology, Wuhan, Hubei 430074, P. R. China}

\begin{abstract}
	Elastic angular momentum is an emerging field, with some controversies on the correct field-theory expressions and the decomposition of longitudinal and transverse components. Motivated by the recent two papers [Phys.Rev.Lett. $\boldsymbol{128}$, 064301(2022), Phys.Rev.Lett. $\boldsymbol{129}$, 204303(2022)] on this issue, we systematically analyze by Noether's theorem the canonical and Belinfante energy-momentem and angular momentum, then explain why the two familiar expressions, together with other various conservered currents, are all correct for elastic wave. Remarkbly, to illustrate the usefullness of different expressions, we give an example on earthquake energy measurement with a new energy density expression which is more advantageous in practical measurement. Moreover, since the elastic wave is distinct from a quantum one, we suggest that the decomposition of longitudinal and transverse components, in fact, makes sense and can be clearly expressed by the observable displacement field. We hope that this paper would clarify the controversies, and finally we give prospect of future work on the Geometric Spin Hall Effect of elastic wave, where the various expressions of conserved currents would exhibit different applications.  
\end{abstract}

	\section{Interest and Controversies Regarding Elastic Angular Momentum}
	In 1992, Allen $\textit{et al}$.\cite{Allen:1992zz} pioneered the work of orbital angular momentum (OAM) of light. Such OAM can be physically realized by the helical wave-front structure with a phase singularity of the light field, and can be extracted by absorptive particles to convert into a mechanical torque \cite{He:1995}. The discovery has attracted much theoretical interest\cite{Barnett:2002,Barnett:2010,Bliokh:2014} and rekindled the interest on optical tweezers\cite{Garce´s-Cha´vez:2003,ChenRui:2020}.\par
	A natural extension is that, elastic wave, as also a classical wave system distinct from electromagnetism, can carry angular momentum. Recently the intrinsic spin of elastic wave has been studied by Long $\textit{et al}.$\cite{Long:2018},  then OAM by Chaplain $\textit{et al}.$\cite{Chaplain:2022}. The authors in \cite{Chaplain:2022} split the displacement field $\boldsymbol{\xi}$ into longitudinal part $\boldsymbol{\xi}_{\parallel}$ and transverse part $\boldsymbol{\xi}_{\perp}$, then showed that the longitudinal part $\boldsymbol{\xi}_{\parallel}$ which was associated with compressional motion, could carry a well-defined orbital angular momentum. Soon afterwards Bliokh\cite{Bliokh:2022} suggested that the theoretical result of Refs. \cite{Chaplain:2022} should be corrected, and claimed that it did not make much sense to separate the longitudinal and transverse parts of the field. \par
	This paper is motivated by the papers \cite{Chaplain:2022,Bliokh:2022} in which disagreements arise on what is the proper definition of elastic wave's energy, momentum and angular momentum and on whether it is possible to single out the compression or the shear field. In Ref.\cite{Chaplain:2022}, the authors described the elastic energy and energy flux by the canonical energy-momentum tensor, which is equation (6) in their paper. Then, the angular momentum density was constructed as $M_{ijk}=x_iT_{jk}-x_jT_{ik}$, which should be the standard procedure for constructing the Belinfante angular momentum, but not the canonical one. And one can check that, the angular momentum charge density $M_{ij0}=\epsilon_{ilm}x_lT_{mj}$ does not conserve. Very soon after Chaplain $\textit{et al}.$, Bliokh\cite{Bliokh:2022} pointed out that, the total angular momentum of the form ``position$\times$energy flux" was not correct, and proposed the canonical angular momentum expression using Noether's theorem with respect to the spatial rotations. Bliokh also suggested that, there was no need to distinguish between the longitudinal, transverse, or mixed contributions to the momentum or the angular momentum, because any probes interacted with the total local displacement field $\boldsymbol{\xi}$ rather than separately the longitudinal field $\boldsymbol{\xi}_{\parallel}$ or the transverse field $\boldsymbol{\xi}_{\perp}$.\par
	One should also notice the difference between angular momentum and pseudo-angular momentum, which correspond to two different types of spatial rotation symmetry (see \cite{Streib:2021} for a nice discussion). In this paper, we focus only on the so-called pseudo-angular momentum (the caononical pseudo-angular momentum corresponds to the conventional rotation transformation $R\boldsymbol{\xi}(R^{-1}\boldsymbol{x})$). For a field-theory system, the pseudo-angular momentum is actually more valuable than the physical one, we will discuss the pseudo-quantity (both momentum and angular momentum) and the physical one in detail in a forthcoming paper.

\par
	To address the controversies, first of all, we give a systematic treatment on the canonical and Belinfante expressions of energy, momentum and angular momentum for elastic wave in section 2. In section 3, we discuss the concrete difference between quantum wave and elastic wave and explain that one can realize local measurement for elastic wave, every physical quantity constructed by the displacement field $\boldsymbol{\xi}$ is observable and the field decomposition makes sense. Next in section 4, we suggest a useful energy expression $T_{\text {new }}^{00}=\frac{1}{2} \rho\left(\dot{\xi}_i \cdot \dot{\xi}_i-\ddot{\xi}_i \cdot \xi_i\right)$, and consider its application in seismology. Finally in section 5, we summarize our discussion and give a prospect on the Geometric Spin Hall Effect of elastic wave.

	\section{Canonical and Belinfante Expressions of Elastic Energy-Momentum and Angular Momentum}
	
	Classical field theory is a powerful tool for investigating energy, momentum and angular momentum of  elastic wave. Noether's theorem\cite{Noether} builds an elegant bridge from continuous symmetries to conservation laws. However, the expression of the conserved charge density with respect to a specific kind of symmetry (the space-time translation invariance, for example) is not unique, one is free to add a total divergence term to a given current density maintaining the charge conservation. In this section, we will recover the result of Refs.\cite{Chaplain:2022,Bliokh:2022}, and illustrate the relation between the canonical angular momentum and the Belinfante angular momentum in elastic wave system. \par

	We start from the Navier-Cauchy Equation which governs the elastic wave:
	\begin{equation}
		\mu \partial_j \partial_j \xi_i+(\lambda+\mu) \partial_j \partial_i \xi_j=\rho \ddot{\xi}_i, 
	\end{equation}
	where and below repeated indices are summed over, Latin indices such as $i$ and $j$ run from 1 to 3. Here $\xi_i$ is the displacement field, $\ddot{\xi}_i$ is the double time derivative of $\xi_i$,  $\lambda$ and $\mu$ are Lam\'e's first and second parameters. One can promptly deduce the Lagrangian density for elastic wave:
	\begin{equation}
		\mathcal{L}=\frac{1}{2} \rho \dot{\xi_i} \cdot \dot{\xi_i}  -\frac{1}{2} \mu\partial_j \xi_i \cdot\partial_j \xi_i
		-\frac{1}{2}(\mu+\lambda)\partial_i \xi_j \cdot\partial_j \xi_i.
	\end{equation} \par

	Applying Noether's theorem, the canonical energy-momentum tensor with respect to space-time translation symmetry is given as:

	\begin{equation}
		T_\text{cano}^{\mu \nu}=\frac{\partial \mathcal{L}}{\partial(\partial_\mu \xi_i)} \partial^\nu \xi_i-g^{\mu \nu} \mathcal{L},
	\end{equation}\\with
	\begin{equation}
		\partial_\mu T_\text{cano }^{\mu \nu}=0,
	\end{equation}
	where $g^{\mu\nu}$ is the Minkowski metric $diag(+,-,-,-)$, Greek indices $\mu$ and $\nu$ run from 0 to 3. The canonical energy density is
	\begin{equation} \label{canoenergyden}
		\begin{aligned}
			T_\text{cano }^{00}&=\frac{\partial \mathcal{L}}{\partial(\dot{\xi_i})} \cdot \dot{\xi_i}-\mathcal{L} \\
			&=\frac{1}{2}[\rho\dot{\xi_i} \cdot \dot{\xi_i}+\mu \partial_j \xi_i \cdot \partial_j \xi_i+(\mu+\lambda) \partial_i \xi_j \cdot \partial_j \xi_i],
		\end{aligned}
	\end{equation}
	and the canonical momentum density is
	\begin{equation}
		\begin{aligned}\label{canomomentumden}
			\mathcal{P}^k_\text{cano}=T_\text{cano}^{0k} & =-\frac{\partial \mathcal{L}}{\partial(\partial\dot{\xi_i})}  \partial_k \xi_i \\
			& =-\rho\dot{\xi_i} \cdot \partial_k \xi_i.
		\end{aligned}
	\end{equation}
	We can represent (\ref{canomomentumden}) in  vector form:
	\begin{equation}\label{canomentumdenvec}
		\boldsymbol{\mathcal{P}}_\text{cano}=-\rho \dot{\boldsymbol{\xi}}\cdot(\boldsymbol{\nabla}) \boldsymbol{\xi}.
	\end{equation}
	Formula (\ref{canoenergyden}) and (\ref{canomentumdenvec}) correspond to equation (2) in Ref.\cite{Bliokh:2022}, with a specific potential energy density $U$. The energy flux is
	\begin{equation}
		\begin{aligned}
			\mathcal{E}^k_\text{cano}=T_\text{cano }^{k0} & =\frac{\partial \mathcal{L}}{\partial(\partial_k \xi_i)} \cdot \dot{\xi_i} \\
			& =-[\mu \partial_k \xi_i+(\lambda+\mu) \partial_i \xi_k] \cdot \dot{\xi_i}, 
		\end{aligned}
	\end{equation}
	or, in vector form:
	\begin{equation}\label{canoenergyflowvec}
		\boldsymbol{\mathcal{E}}_\text{cano}=-\mu \dot{\boldsymbol{\xi}}\cdot(\boldsymbol{\nabla}) \boldsymbol{\xi}-(\lambda+\mu)(\dot{\boldsymbol{\xi}}\cdot\boldsymbol{\nabla}) \boldsymbol{\xi}.
	\end{equation}
	Compared with (\ref{canomentumdenvec}), the extra part $(\dot{\boldsymbol{\xi}}\cdot\boldsymbol{\nabla}) \boldsymbol{\xi}$ in canonical energy flux indicates an intrinsic screwing structure. The canonical momentum flux reads
	\begin{equation}
		\begin{aligned}
			T_\text{cano}^{kj} & =\frac{\partial \mathcal{L}}{\partial(\partial_k \xi_i)} \cdot \partial^j \xi_i-g^{k j} \mathcal{L} \\
			& =\left[\mu \partial_k \xi_i+(\mu+\lambda) \partial_i \xi_k\right] \cdot \partial_j \xi_i-g^{k j} \mathcal{L}.
		\end{aligned}
	\end{equation}\par
	As mentioned above, the energy-momentum tensor is not unique, and we will argue in the next section that, the different types of energy-momentum tensor in elastic system are all correct but contain distinct information. In what follows we will consider the conventional  Belinfante energy-momentum tensor\cite{Belinfante:1939}: 
	\begin{equation}
		T_\text{Bel}^{\mu \nu}=T_\text{cano}^{\mu \nu}+\frac{1}{2} \partial_\rho [S^{\rho \mu \nu}+S^{\mu  \nu \rho}+S^{ \nu \mu \rho}],
	\end{equation}\\with
	\begin{equation}
		\partial_\mu T_\text{Bel}^{\mu \nu}=0.	
	\end{equation}
	Here, $S^{\rho \mu\nu }$ is the spin ``tensor" defined as:
	\begin{equation}
		S^{\rho  \mu \nu}=\textrm{i}\frac{\partial \mathcal{L}}{\partial(\partial_\rho \xi_i)} \Sigma_{i j}^{\mu \nu} \xi_j,
	\end{equation}
and $\Sigma_{i j}^{\mu \nu}$ is the spin matrix
\begin{equation}
\Sigma_{i j}^{\mu \nu}=\textrm{i}(\delta^{\mu}_{i}\delta^{\nu}_{j}-\delta^{\nu}_{i}\delta^{\mu}_{j}).
\end{equation} \par
	 Notice that the displacement $\xi_i$ is a three-dimensional vector field and has no time component compared with a relativistic field, thus, such Belinfante tensor can not be completely a symmetric tensor and it is only symmetric with the spatial indices $k$ and $j$. Following similar steps, we have the Belinfante energy density as

	\begin{equation}\label{canenergyequalBel}
		T_\text{Bel}^{00}=T_\text{cano }^{00}=\frac{1}{2}[\rho\dot{\xi_i}  \cdot \dot{\xi_i}+\mu \partial_j \xi_i \cdot \partial_j \xi_i+(\mu+\lambda) \partial_i \xi_j \cdot \partial_j \xi_i],
	\end{equation}
	and the Belinfante momentum density as
	\begin{equation}\label{Belmomentumden}
		\mathcal{P}_\text{Bel}^k=T_\text{Bel }^{0 k}=-\rho \dot{\xi_i} \cdot \partial_k \xi_i+\frac{1}{2} \rho \partial_i[\dot{\xi_i}  \xi_k-\dot{\xi_k}  \xi_i].
	\end{equation}
	We can cast (\ref{Belmomentumden}) into vector form:
	\begin{equation}\label{Belmomentumdenvec}
		\boldsymbol{\mathcal{P}}_\text{Bel}=-\rho \dot{\boldsymbol{\xi}}\cdot(\boldsymbol{\nabla}) \boldsymbol{\xi}+\frac{1}{2}\rho \boldsymbol{\nabla}\cdot \left[ \dot{\boldsymbol{\xi}}(\boldsymbol{\xi})-(\dot{\boldsymbol{\xi}})\boldsymbol{\xi}\right].
	\end{equation}
	The Belinfante energy flux reads:
	\begin{equation}
		{\mathcal{E}_\text{Bel}^{k}=T_\text{Bel }^{k0}=-\mu \partial_k \xi_i \cdot \dot{\xi_i}-(\mu+\lambda) \partial_i \xi_k \cdot \dot{\xi_i}-\frac{1}{2} \rho\partial_i[\dot{\xi_k} \cdot \xi_i-\dot{\xi_i} \cdot \xi_k],}
	\end{equation}
	or in vector form:
	\begin{equation}\label{Belenergyflowvec}
		\boldsymbol{\mathcal{E}}_\text{Bel}=-\mu \dot{\boldsymbol{\xi}}\cdot(\boldsymbol{\nabla}) \boldsymbol{\xi}-(\mu+\lambda)(\dot{\boldsymbol{\xi}}\cdot\boldsymbol{\nabla}) \boldsymbol{\xi}+\frac{1}{2}\rho \boldsymbol{\nabla}\cdot \left[ \dot{\boldsymbol{\xi}}(\boldsymbol{\xi})-(\dot{\boldsymbol{\xi}})\boldsymbol{\xi}\right].
	\end{equation}
	The Belinfante momentum flux is:
	\begin{equation}\label{Belmomentumfluxden}
		\begin{aligned}
			T_\text{Bel}^{kj}= & \mu\left[\partial_k \xi_i \cdot \partial_j \xi_i-\partial_i \xi_k \cdot \partial_i \xi_j+\partial_k \xi_j \cdot \partial_i \xi_i+\partial_j \xi_k \cdot \partial_i \xi_i\right] \\
			&+\frac{1}{2} \mu\left[2\partial_i \partial_k \xi_j \cdot \xi_i+2\partial_i \partial_j \xi_k \cdot \xi_i-\partial_i \partial_i \xi_j \cdot \xi_k-\partial_i \partial_i \xi_k \cdot \xi_j\right. \\
			& \left.-\partial_k \partial_i \xi_i \cdot \xi_j-\partial_j \partial_i \xi_i \cdot \xi_k\right] \\
			&+\frac{1}{2}\lambda \left[ \partial_k \xi_i \cdot \partial_i \xi_j+\partial_j \xi_k \cdot \partial_i \xi_i+\partial_k \xi_j \cdot \partial_i \xi_i+ \partial_i \xi_k \cdot \partial_j \xi_i\right. \\
			& \left.-2\partial_i \xi_j \cdot \partial_i \xi_k\right] \\
			&+ \frac{1}{2}\lambda \left[\partial_i \partial_j \xi_k \cdot \xi_i-\partial_i \partial_i \xi_k \cdot \xi_j+\partial_i \partial_k \xi_j \cdot \xi_i-\partial_i \partial_i \xi_j \cdot \xi_k\right] \\
			& -g^{k j}\mathcal{L},
		\end{aligned}
	\end{equation}
	which is symmetric under the exchange of $k$ and $j$.\par
	We continue to consider the canonical angular momentum under the invariance of spatial rotations: 
	\begin{equation} \label{CanoAM1}
		M_{\text {cano }}^{\mu \alpha \beta}=x^\alpha T_{\text {cano }}^{\mu \beta}-x^\beta T_{\text {cano }}^{\mu \alpha}+\textrm{i}\frac{\partial \mathcal{L}}{\partial\left(\partial_\mu \xi_i\right)} \Sigma_{i j}^{ \alpha\beta}\xi_j,
	\end{equation}
	with
	\begin{equation}
		\partial_\mu M_{\text {cano}}^{\mu m n}=0.
	\end{equation}
	The canonical angular momentum density is
	\begin{equation}
		\mathcal{J}_{\text {cano }}^{m n}=M_{\text {cano }}^{0 m n}=\rho \dot{\xi}_i \left(x^m \partial^n \xi_i-x^n \partial^m \xi_i\right)+\rho(\dot{\xi}^n \xi^m-\dot{\xi}^m \xi^n),
	\end{equation}
	and one can also put it into vector form:
	
	\begin{equation}\label{canoangularmomentumdenvec}
		\boldsymbol{\mathcal{J}}_\text{cano}=-\rho \dot{\boldsymbol{\xi}}\cdot(\boldsymbol{r}\times\boldsymbol{\nabla}) \boldsymbol{\xi}-\rho\dot{\boldsymbol{\xi}}\times \boldsymbol{\xi},
	\end{equation}
	which is the equation (3) in \cite{Bliokh:2022}.

One should note that the orbital part in (\ref{CanoAM1}) constructed with the Canonical energy-momentum tensor is not conserved separately. If one would like to construct a conserved angular momentum in an ``orbital" form, one should use the Belinfante energy-momentum tensor
	\begin{equation}
		M_{\text {Bel }}^{\mu \alpha \beta}=x^\alpha T_{\text {Bel}}^{\mu \beta}-x^\beta T_{\text {Bel}}^{\mu \alpha},
	\end{equation}
	with
	\begin{equation}
		\partial_\mu M_{\text {Bel}}^{\mu m n}=0.
	\end{equation}
	The Belinfante angular momentum density reads:
	\begin{equation}
		\begin{aligned}
			\mathcal{J}_{\text {Bel}}^{m n}=M_{\text {Bel}}^{0mn }=&\rho \dot{\xi}_i\left(x^m \partial^n \xi_i-x^n \partial^m \xi_i\right) \\
			& -\frac{1}{2} \rho\left[x^m \partial_i\left(\dot{\xi}_i \xi^n-\dot{\xi}^n \xi_i\right)-x^n \partial_i\left(\dot{\xi}_i \xi^m-\dot{\xi}^m \xi_i\right)\right],
		\end{aligned}
	\end{equation}
	or in vector form:
	\begin{equation}\label{Belangularmomentumdenvec}
		\boldsymbol{\mathcal{J}}_\text{Bel}=-\rho \dot{\boldsymbol{\xi}}\cdot(\boldsymbol{r}\times\boldsymbol{\nabla}) \boldsymbol{\xi}-\frac{1}{2}\rho\left\{\boldsymbol{r}\times\left[\boldsymbol{\nabla}\cdot[(\boldsymbol{\xi})\dot{\boldsymbol{\xi}}-\boldsymbol{\xi}(\dot{\boldsymbol{\xi}})]\right] \right\}.
	\end{equation}\par
	Next, we will go on to show that  provided that one is allowed to drop the surface term, which means the field vanishes at spatial infinity, the charge of the canonical or the Belinfante-type tensor  is the same.\par
	For the momentum, one has
	\begin{equation}
		\begin{aligned}
			\boldsymbol{\text{P}}&=\int\text{d}^3x\boldsymbol{\mathcal{P}}_\text{cano}\\&=\int\text{d}^3x\boldsymbol{\mathcal{P}}_\text{Bel}-\frac{1}{2}\rho\int\text{d}^3x \boldsymbol{\nabla}\cdot \left[ \dot{\boldsymbol{\xi}}(\boldsymbol{\xi})-(\dot{\boldsymbol{\xi}})\boldsymbol{\xi}\right]\\&=\int\text{d}^3x\boldsymbol{\mathcal{P}}_\text{Bel},
		\end{aligned}
	\end{equation}
	as long as the quantity ($\dot{\xi_i}  \xi_k-\dot{\xi_k}  \xi_i$) vanishes at the boundary.\par
	For the angular momentum, one has
	\begin{equation}
		\begin{aligned}
			\boldsymbol{\text{J}}&=\int\text{d}^3x\boldsymbol{\mathcal{J}}_\text{cano}\\&=\int\text{d}^3x\boldsymbol{\mathcal{J}}_\text{Bel}-\frac{1}{2}\rho\int\text{d}^3x \boldsymbol{\nabla}\cdot \left\lbrace  \boldsymbol{r}\times[\dot{\boldsymbol{\xi}}(\boldsymbol{\xi})-(\dot{\boldsymbol{\xi}})\boldsymbol{\xi}]\right\rbrace \\&=\int\text{d}^3x\boldsymbol{\mathcal{J}}_\text{Bel},
		\end{aligned}
	\end{equation}
	as long as the surface term
	$\left[x^m\left(\dot{\xi}_i \xi^n-\dot{\xi}^n \xi_i\right)\right. \left.-x^n\left(\dot{\xi}_i \xi^m-\dot{\xi}^m \xi_i\right)\right]$ vanishes at the boundary.\par

	\par Although the momentum density, momentum flux and energy flux expressions of the canonical and the Belinfante construction seem quite distinct,  we are to point out that, these expressions are all observable and have different uses.

	\section{Decomposition Into Longitudinal and Transverse Components: Elastic Wave and Quantum Wave}
	For elastic wave, the basic observable is displacement field $\boldsymbol{\xi}$. Suppose one puts sensing elements in the media to measure the displacement, provided that the scale of the sensors are very small compared with the wave length, one is possible to realize local measurements. In principle, such measurements would not obviously alter the properties of the wave system, since the wave is completely classical. In Ref.\cite{Bliokh:2022}, Bliokh clamed that it did not make much sense to separate
	the longitudinal, transverse, and hybrid contributions to elstic wave's momentum and angular momentum. To the contrary, we suggest that, every quantity $f(\boldsymbol{\xi})$ constructed by $\boldsymbol{\xi}$ makes sense and the field decomposition is effective. First, the displacement can be separated into  longitudinal and transverse components
	\begin{equation} \label{longtransdecomposition}
		\boldsymbol{\xi}=\boldsymbol{\xi}_{\parallel}+\boldsymbol{\xi}_{\perp},
	\end{equation}
	where $\boldsymbol{\xi}_{\parallel}$ is the longitudinal field and satisfies $\boldsymbol{\nabla}\times\boldsymbol{\xi}_{\parallel}=0$, $\boldsymbol{\xi}_{\perp}$ is the transverse field and satisfies $\boldsymbol{\nabla}\cdot\boldsymbol{\xi}_{\perp}=0$.

	We can represent $\boldsymbol{\xi}_{\perp}$ and $\boldsymbol{\xi}_{\parallel}$ by $\boldsymbol{\xi}$\cite{Chen:2009}:
	\begin{equation}
		\boldsymbol{\xi}_{\parallel}=\boldsymbol{\nabla}\frac{1}{\boldsymbol{\nabla}^{2}}(\boldsymbol{\nabla} \cdot \boldsymbol{\xi}),
	\end{equation}
	\begin{equation}
		\boldsymbol{\xi}_{\perp}=-\boldsymbol{\nabla}\times\frac{1}{\boldsymbol{\nabla}^{2}}(\boldsymbol{\nabla}\times\-\boldsymbol{\xi})=\boldsymbol{\xi}-\boldsymbol{\nabla}\frac{1}{\boldsymbol{\nabla}^{2}}(\boldsymbol{\nabla} \cdot \boldsymbol{\xi}).
	\end{equation}
It should be noted that $\frac{1}{\boldsymbol{\nabla}^{2}}$ is an integral operator, e.g., if $\boldsymbol{\nabla}^2 g=\boldsymbol{\nabla} \cdot \boldsymbol{\xi}$, then
	\begin{equation}
g=\frac{1}{\boldsymbol{\nabla}^{2}}(\boldsymbol{\nabla} \cdot \boldsymbol{\xi})\equiv-\frac{1}{4\pi}\int\text{d}^3 x^{\prime}\frac{\boldsymbol{\nabla} \cdot \boldsymbol{\xi}}{\left|\boldsymbol{x}-\boldsymbol{x}^{\prime} \right| }.
\end{equation}
We have required that, for a finite system, $g$ vanishes at infinity.

	Let us take the canonical momentum density for example. Substituting (\ref{longtransdecomposition}) into (\ref{canomentumdenvec}), one has 
	\begin{equation}\label{canomentumdenvecdecomposi}
		\boldsymbol{\mathcal{P}}_\text{cano}=\boldsymbol{\mathcal{P}}_{\parallel}+\boldsymbol{\mathcal{P}}_{\perp}+\boldsymbol{\mathcal{P}}_\text{mix},
	\end{equation}
	with
	\begin{equation}
		\boldsymbol{\mathcal{P}}_{\parallel}=-\rho \dot{\boldsymbol{\xi}}_{\parallel}\cdot(\boldsymbol{\nabla}) \boldsymbol{\xi}_{\parallel},
	\end{equation}
	\begin{equation}
		\boldsymbol{\mathcal{P}}_{\perp}=-\rho \dot{\boldsymbol{\xi}}_{\perp}\cdot(\boldsymbol{\nabla}) \boldsymbol{\xi}_{\perp},
	\end{equation}
	and
	\begin{equation}
		\boldsymbol{\mathcal{P}}_\text{mix}=-\rho\left[  \dot{\boldsymbol{\xi}}_{\parallel}\cdot(\boldsymbol{\nabla}) \boldsymbol{\xi}_{\perp}+\dot{\boldsymbol{\xi}}_{\perp}\cdot(\boldsymbol{\nabla}) \boldsymbol{\xi}_{\parallel}\right],
	\end{equation}
	where $	\boldsymbol{\mathcal{P}}_{\parallel}$ is the longitudinal contribution to the momentum density, $	\boldsymbol{\mathcal{P}}_{\perp}$ the transverse contribution and $\boldsymbol{\mathcal{P}}_\text{mix}$ the hybrid contribution.
	\par We can also present $\boldsymbol{\mathcal{P}}_{\parallel}$, $\boldsymbol{\mathcal{P}}_{\perp}$ and $	\boldsymbol{\mathcal{P}}_\text{mix}$ with $\boldsymbol{\xi}$:
	\begin{equation}
		\boldsymbol{\mathcal{P}}_{\parallel}=-\rho \boldsymbol{\nabla}\frac{\boldsymbol{\nabla} \cdot \dot{\boldsymbol{\xi}}}{\boldsymbol{\nabla}^{2}}\cdot(\boldsymbol{\nabla}) \boldsymbol{\nabla}\frac{\boldsymbol{\nabla} \cdot \boldsymbol{\xi}}{\boldsymbol{\nabla}^{2}},
	\end{equation}
	\begin{equation}
		\boldsymbol{\mathcal{P}}_{\perp}=-\rho\left[  \dot{\boldsymbol{\xi}}\cdot(\boldsymbol{\nabla}) \boldsymbol{\xi}-\dot{\boldsymbol{\xi}}\cdot(\boldsymbol{\nabla}) \boldsymbol{\nabla}\frac{\boldsymbol{\nabla} \cdot \boldsymbol{\xi}}{\boldsymbol{\nabla}^{2}}-\boldsymbol{\nabla}\frac{\boldsymbol{\nabla} \cdot \dot{\boldsymbol{\xi}}}{\boldsymbol{\nabla}^{2}}\cdot(\boldsymbol{\nabla})  \boldsymbol{\xi}+\boldsymbol{\nabla}\frac{\boldsymbol{\nabla} \cdot \dot{\boldsymbol{\xi}}}{\boldsymbol{\nabla}^{2}}\cdot(\boldsymbol{\nabla}) \boldsymbol{\nabla}\frac{\boldsymbol{\nabla} \cdot \boldsymbol{\xi}}{\boldsymbol{\nabla}^{2}}\right] ,
	\end{equation}
	and
	\begin{equation}
		\boldsymbol{\mathcal{P}}_\text{mix}=-\rho\left[  \boldsymbol{\nabla}\frac{\boldsymbol{\nabla} \cdot \dot{\boldsymbol{\xi}}}{\boldsymbol{\nabla}^{2}}\cdot(\boldsymbol{\nabla}) \boldsymbol{\xi}-\boldsymbol{\nabla}\frac{\boldsymbol{\nabla} \cdot \dot{\boldsymbol{\xi}}}{\boldsymbol{\nabla}^{2}}\cdot(\boldsymbol{\nabla}) \boldsymbol{\nabla}\frac{\boldsymbol{\nabla} \cdot \boldsymbol{\xi}}{\boldsymbol{\nabla}^{2}}+\dot{\boldsymbol{\xi}}\cdot(\boldsymbol{\nabla}) \boldsymbol{\nabla}\frac{\boldsymbol{\nabla} \cdot \boldsymbol{\xi}}{\boldsymbol{\nabla}^{2}}-\boldsymbol{\nabla}\frac{\boldsymbol{\nabla} \cdot \dot{\boldsymbol{\xi}}}{\boldsymbol{\nabla}^{2}}\cdot(\boldsymbol{\nabla}) \boldsymbol{\nabla}\frac{\boldsymbol{\nabla} \cdot \boldsymbol{\xi}}{\boldsymbol{\nabla}^{2}}\right].
	\end{equation}
	Thus, these contributions are all functions of observable $\boldsymbol{\xi}$ and can be single out explicitly.\par
		For a quantum system, the wave function $\psi(\boldsymbol{x},t)$ is not an observable. Distinct from the local measurement of displacement $\boldsymbol{\xi}$, in the quantum regime it is not possible to strictly realize a  local measurement. In case of a quantum measurement, to detect a photon for example, one catches the energy, momentum and angular momentum of the whole wave at a single point-like tiny region. The concept of energy (momentum or angular momentum) in the wave description is an integral of the energy (momentum or angular momentum) density. As a result, it is not possible to single out a longitudinal or transverse field contribution in a local sense.   
		\par
	An interesting example is the decomposition of the electric field $\boldsymbol{E}$:
	\begin{equation}
		\boldsymbol{E}=\boldsymbol{E}_{\parallel}+\boldsymbol{E}_{\perp},
	\end{equation}
where $\boldsymbol{E}_{\parallel}$ is the longitudinal field and satisfies $\boldsymbol{\nabla}\times\boldsymbol{E}_{\parallel}=0$, $\boldsymbol{E}_{\perp}$ is the transverse field and satisfies $\boldsymbol{\nabla}\cdot\boldsymbol{E}_{\perp}=0$. When $\boldsymbol{E}$ stands for classical electric field which is an observable just similar to the displacement field $\boldsymbol{\xi}$, it is meaningful to discuss the longitudinal or the transverse contribution as a local field. In contrast, when $\boldsymbol{E}$ stands for photon field in the quantum electrodynamics, it should be quantized into creation and annihilation operators and the observable is S-matrix. It is not practical to split the longitudinal or the transverse contribution since the field itself is not an observable.\par 
One can go a step further to consider the decomposition of gauge field $\boldsymbol{A}$:
	\begin{equation}
		\boldsymbol{A}=\boldsymbol{A}_{\parallel}+\boldsymbol{A}_{\perp},
	\end{equation}
with $\boldsymbol{\nabla}\times\boldsymbol{A}_{\parallel}=0$ and $\boldsymbol{\nabla}\cdot\boldsymbol{A}_{\perp}=0$. Even in classical field theory, the gauge field $\boldsymbol{A}$ is not a observable and can not be measured locally. Thus, in the sense of local measurement, it is meaningless to decompose $\boldsymbol{A}$\cite{Leader:2014}.
	
\section{Usefulness of Different Energy Expressions: An Earthquake Wave Example}
	In this section we draw attention to the different energy density expressions, and show that there can be a remarkably useful expression for calculating the total energy of the system. First recall that in equation (\ref{canenergyequalBel}), the canonical energy density is equal to the Belinfante energy density. And the energy density is of the form $\mathcal{H}=\mathcal{T}+\mathcal{V}$, where $\mathcal{T}=\frac{1}{2}\rho\dot{\xi_i}  \cdot \dot{\xi_i}$ denotes the ``kinetic energy" and $\mathcal{V}=\frac{1}{2}\mu \partial_j \xi_i \cdot \partial_j \xi_i+\frac{1}{2}(\mu+\lambda) \partial_i \xi_j \cdot \partial_j \xi_i$ denotes the ``potential energy". Thus, it truly describes the energy that the elastic wave carries. One can take a closer look to find that, the energy density
	expression (\ref{canenergyequalBel}) contains not only time derivatives but also spatial derivatives of $\boldsymbol{\xi}$. \par
\begin{figure}[!h]
		\centering
		\includegraphics[width=0.52\textwidth]{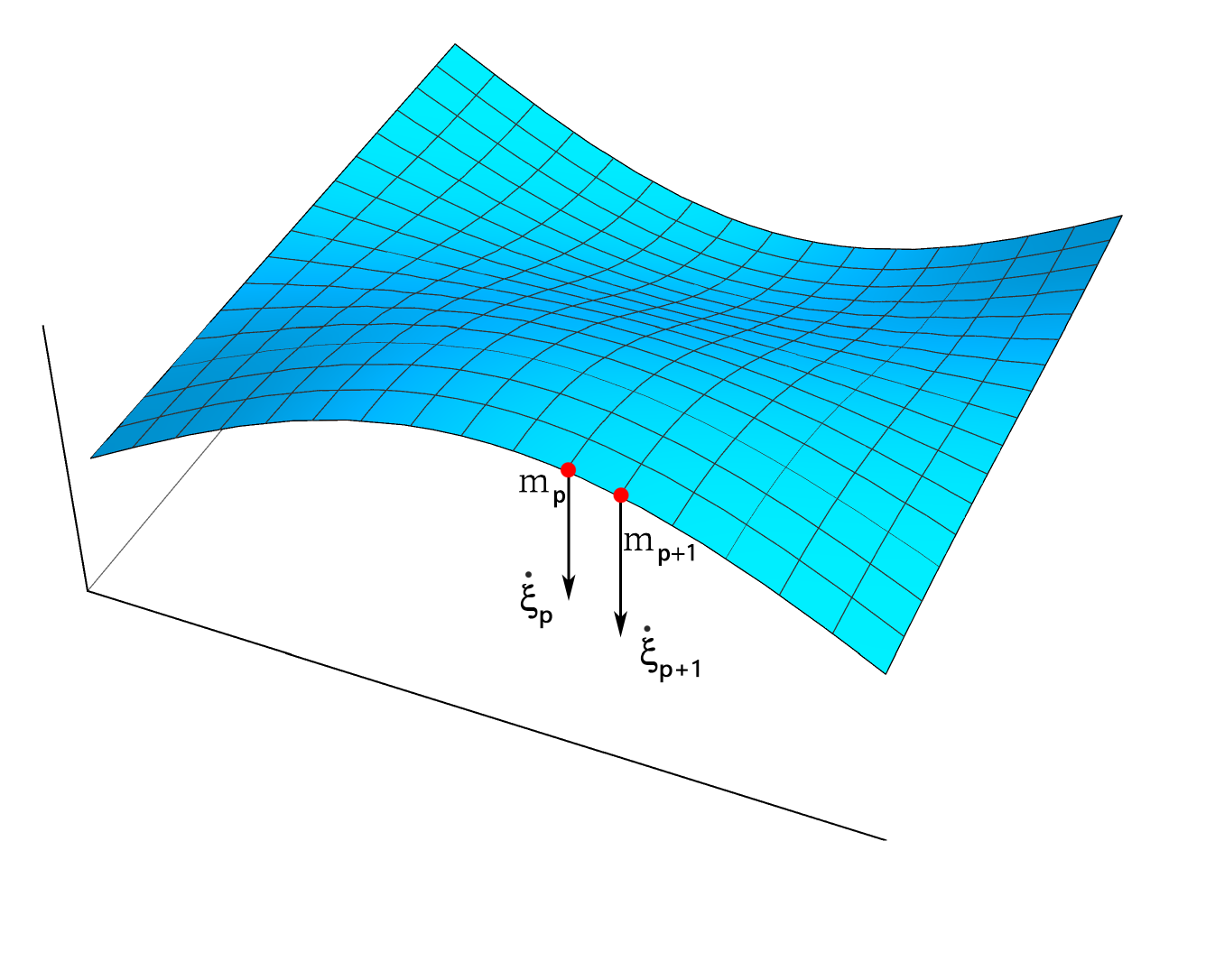}
		\caption{\textit{A sketch of a waving surface. The two red dots represent two adjacent displacement sensors that being fixed at two nearby mass elements.}} 
		\label{aWavingSurface}
	\end{figure}

	An illustrative example is the energy measurement of earthquakes. Figure \ref{aWavingSurface}	
	is a sketch of a waving surface at a given time, and the surface has been divided into many little parts by the grid lines. To test the total wave energy, one places  in each intersection of the grid lines a sensor for displacement measurement, and finally gathers all the information from these sensors. Suppose now we have two nearby mass elements $m_p$ and $m_{p+1}$, which are equipped with displacement sensors, according to formula (\ref{canenergyequalBel}), the velocity of the elements $\dot{\boldsymbol{\xi}}_p$, $\dot{\boldsymbol{\xi}}_{p+1}$ should be measured, 
	and the adjacent displacements $\boldsymbol{\xi}_p$, $\boldsymbol{\xi}_{p+1}$ are also needed for calculating the spatial derivatives of $\boldsymbol{\xi}$. In the following, we are going to show that, there can be a more efficient energy density expression which removes the spatial
	correlations of different elements when summing up for the total energy.\par
	Motivated by some consideration on quantum measurement, Wang $\textit{et al}$.\cite{Chen:2017} proposed a new energy-momentum tensor. When applying the new expression to the elasticity case, one has 
	\begin{equation}
		T_\text{new}^{\mu \nu}=\frac{\partial \mathcal{L}}{\partial\left(\partial_\mu \xi_i\right)} \overset{\leftrightarrow}{\partial}{}^\nu \xi_i
	\end{equation}
and
	\begin{equation}
	\partial_\mu T_\text{new}^{\mu \nu}=0,	
\end{equation}
where
	\begin{equation}
		\overset{\leftrightarrow}{\partial}{}^{\nu}=\frac{1}{2}(\overset{\rightarrow}{\partial}{}^\nu-\overset{\leftarrow}{\partial}{}^\nu). 
	\end{equation}
	Specially, the new energy density expression  reads:
	\begin{equation} \label{newenergydensityEx}
		T_{\text {new }}^{00}=\frac{1}{2} \rho\left(\dot{\xi}_i \cdot \dot{\xi}_i-\ddot{\xi}_i \cdot \xi_i\right).
	\end{equation}
This is a notable result since it is free of spatial derivatives compare to $T_\text{cano}^{00}$. As a consequence, the measurement of total energy can be done by directly counting up the local contribution of each sensor (i.e., every sensor provides its displacement, velocity and acceleration data), without carefully considering the differences of spatially adjacent contributions. We believe that the new expression (\ref{newenergydensityEx}) will give hints on the high-precision energy measurement in seismology. 
\par
As a comparison, the new type of momentum density is
\begin{equation}
	T_{\text {new }}^{0 m}=\frac{1}{2} \rho\left(\partial_m \dot{\xi}_i \cdot \xi_i-\partial_m \xi_i \cdot \dot{\xi}_i\right),
\end{equation}
or, in vector form
\begin{equation}\label{newmentumdenvec}
	\boldsymbol{\mathcal{P}}_\text{new}=-\frac{1}{2}\rho \dot{\boldsymbol{\xi}}\cdot(\boldsymbol{\nabla}) \boldsymbol{\xi}+\frac{1}{2}\rho\boldsymbol{\xi}\cdot(\boldsymbol{\nabla}) \dot{\boldsymbol{\xi}}.
\end{equation}
We can also put forward the new angular momentum tensor:
	\begin{equation}
		M_{\text {new }}^{\mu \alpha \beta}=x^\alpha T_{\text {new }}^{\mu \beta}-x^\beta T_{\text {new }}^{\mu \alpha}+S^{\mu  \alpha \beta}+\frac{1}{2}\left(g^{\mu \beta} \frac{\partial \mathcal{L}}{\partial\left(\partial_\alpha \xi_i\right)}-g^{\mu \alpha} \frac{\partial \mathcal{L}}{\partial\left(\partial_\beta \xi_i\right)}\right) \xi_i,
	\end{equation}
with
\begin{equation}
	\partial_\mu M_{\text {new}}^{\mu m n}=0.
\end{equation}
The corresponding angular momentum density is
	\begin{equation}
		\begin{aligned}
			M_{\text {new }}^{0 m n}=& x^m T_{\text {new }}^{0 n}-x^n T_{\text {new }}^{0 m}+S^{0 mn} \\	= & \frac{\rho}{2} x^m\left(\partial_n \dot{\xi}_i \cdot \xi_i-\partial_n \xi_i \cdot \dot{\xi}_i\right)-\frac{\rho}{2} x^n\left(\partial_m \dot{\xi}_i \cdot \xi_i-\partial_m \xi_i \cdot \dot{\xi}_i\right)  -\rho\left(\dot{\xi}^m \xi^n-\xi^m \dot{\xi}^n\right),
		\end{aligned}
	\end{equation}
	or, in vector form
		\begin{equation}\label{newangularmomentumdenvec}
		\boldsymbol{\mathcal{J}}_\text{new}=-\frac{1}{2}\rho \dot{\boldsymbol{\xi}}\cdot(\boldsymbol{r}\times\boldsymbol{\nabla}) \boldsymbol{\xi}+\frac{1}{2}\rho \boldsymbol{\xi}\cdot(\boldsymbol{r}\times\boldsymbol{\nabla}) \dot{\boldsymbol{\xi}}-\rho\dot{\boldsymbol{\xi}}\times \boldsymbol{\xi}.
	\end{equation}
	The new angular momentum density expression $\boldsymbol{\mathcal{J}}_\text{new}$ inevitably contains both spatial and time derivatives, just as $\boldsymbol{\mathcal{J}}_\text{cano}$ and $\boldsymbol{\mathcal{J}}_\text{Bel}$.

	\section{Summary and Prospect}
We have visited the current status of discussions on the appropriate expressions of elastic angular momentum, and decomposition of displacement field into transverse and longitudinal part. Guided by the space-time translation symmetry, the canonical energy-momentum of elastic wave is given, then, the Belinfante energy-mementum is also given by the conventional Belinfante symmetrization. For elastic wave, the observation is directly the displacement field $\boldsymbol{\xi}$, so physical quantities that calculated by $\boldsymbol{\xi}$ can all be regarded as observables, including the canonical energy-momentum and the Belinfante one. We therefore conclude that the two conserved currents are both correct and have importance of informatics.
For the same reason, contrary to Bliokh\cite{Bliokh:2022}, we suggest that the field decomposition also makes sense, and the transverse or longitudinal field has been expressed by the original observable $\boldsymbol{\xi}$. The above features of elastic wave is quite different form  quantum wave, where the wave function $\psi(\boldsymbol{x},t)$ is not an observable and the field decomposition does not make much sense.
\par
Since the conserved energy-momentum is not unique, we can explore the variations of  energy-momentum expressions with significant importance. For example, a new type of expression with only local time derivatives is given to offer an alternative method calculating the total energy of earthquake. 
 The different expressions of energy, momentum and angular momentum would have potential applications on the analysis of earthquake, especially for extracting new information which we do not notice before among the massive data. 
 
 \par A very interesting extension is to investigate the Geometry Spin Hall Effect (GSHE) of elastic wave, since all the conserved currents of elastic system are 
 observables. The GSHE of light is a pure geometry-induced shift in the centroid of light intensity which is caused by the spin angular momentum, and similar Hall effect would also be caused by orbital angular momentum. It was discovered theoretically in 2009 by Aiello $\textit{et al}$.\cite{Aiello:2009}, and then experimentally confirmed in 2014 by  Korger $\textit{et al}$.\cite{Korger:2014}. Discovery of the GSHE of light has attracted much interest, to list just a few\cite{Neugebauer:2014,Ling:2017,Bliokh:2019,Wang:2019}. This effect says if a circularly polarized light beam is detected by a screen not perpendicular to the propagating direction, a shift of the beam's centroid occurs by
 \begin{equation} \label{GSHELconven}
 	\langle  y \rangle_P\approx \frac{\lambda}{4\pi}\sigma\tan{\theta},
 \end{equation}
 where $\lambda$ is the wave length, $\sigma=\pm1$ the helicity and $\theta$ the tiltled angle of the screen. In paper\cite{Wang:2019}, Wang  and Chen propose a very convenient technique to calculate this shift by applying a sum rule of the angular momentum tensor. They pointed out that, the result (\ref{GSHELconven}) corresponded to the torque of the Poynting vector (the energy flux of the Belinfante energy-momentum tensor), but when evaluating the light intensity by the Belinfante momentum flux, one had
 \begin{equation} \label{GSHELBmomen}
 	\langle  y \rangle_T\approx \frac{\lambda}{2\pi}\sigma\tan{\theta},
 \end{equation}
 which differed from the conventional GSHE of light by a factor 2. The result presents an important guidance for our future work on the GSHE of elastic wave. In elastic system, the same shift on the centroid of the wave would occur, and calculating the shift with different conserved current would also give different results. In particular, since the various conserved currents of elastic wave can all be regarded as observables, the different expressions are of practical uses.
 \par

	\section*{Acknowledgments}
	This work is supported by the China NSF via Grants No. 11535005.


		\end{document}